\documentclass[11pt]{article}

\pdfoutput=1

\usepackage[margin=20mm]{geometry}

\usepackage{xspace}
\usepackage{paralist}
\usepackage{amsbsy,amssymb,amsmath,amsfonts}
\usepackage{fancybox}
\usepackage{color}
\usepackage{graphicx}
\usepackage{wrapfig}
\usepackage{url}
\usepackage{subfig}
\usepackage{verbatim}

\usepackage{multirow}

\usepackage{algorithm}
\usepackage{algorithmic}

\usepackage{multibib}
\newcites{app}{Additional References}

\DeclareGraphicsExtensions{.png}
\DeclareGraphicsExtensions{.jpg}
\newcommand{\ignore}[1]{}

\newcommand{\Z}{\mathbb{Z}}

\newcommand{\N}{\mathbb{N}}
\newcommand{\RR}{\mathbb{R}}
\newcommand{\C}{\mathbb{C}}

\newcommand{\cgal}{\textsc{Cgal}\xspace}

\newcommand{\gmp}{\textsc{Gmp}\xspace}
\newcommand{\ntl}{\textsc{Ntl}\xspace}

\newcommand{\bs}{\textsc{Bisolve}\xspace}
\newcommand{\rs}{\textsc{Rs}\xspace}
\newcommand{\isolate}{\textsc{Isolate}\xspace}
\newcommand{\lgp}{\textsc{Lgp}\xspace}

\newcommand{\etal}{et~al.\xspace}

\def\marrow{\marginpar[\hfill$\longrightarrow$]{$\longleftarrow$}}
\def\michael#1{\textcolor{red}{\textsc{Michael says: }{\marrow\sf #1}}}
\def\eric#1{\textcolor{red}{\textsc{Eric says: }{\marrow\sf #1}}}
\def\pavel#1{\textcolor{red}{\textsc{Pavel says: }{\marrow\sf #1}}}
\def\lncs#1{\textcolor{red}{\textsc{LNCS says: }{\marrow\sf #1}}}

\newif\ifwithexperiments
\withexperimentsfalse

\numberwithin{equation}{section}
\numberwithin{figure}{section}

\usepackage{amssymb,amsmath}
\usepackage{graphicx}
\usepackage{rotating}

\usepackage{caption}

\usepackage{url}
\urldef{\mails}\path|{eric,asm,msagralo}@mpi-inf.mpg.de|

\newtheorem{theorem}{Theorem}
\newtheorem{lemma}{Lemma}

\providecommand{\qed}{\rule[-0.2ex]{0.3em}{1.4ex}}

\newlength{\proofpostskipamount}\newlength{\proofpreskipamount}
\newenvironment{proof}%
               {\par\vspace{\proofpreskipamount}\noindent{\bf Proof:}\hspace{0.5em}}% 0.5 before
               {\nopagebreak%
                \strut\nopagebreak%
                \hspace{\fill}\qed\par\vspace{\proofpostskipamount}\noindent}

\newcommand{\bdesc}{$\textsc{Bdc}$~}

\begin{document}

% \mainmatter  % start of an individual contribution

% first the title is needed
\title{An Elimination Method for Solving Bivariate Polynomial Systems: Eliminating the Usual Drawbacks}

% a short form should be given in case it is too long for the running head
% \titlerunning{Bivariate Real Root Solving - with a twist}

% the name(s) of the author(s) follow(s) next
%
% NB: Chinese authors should write their first names(s) in front of
% their surnames. This ensures that the names appear correctly in
% the running heads and the author index.
%

% \numberofauthors{3} 
% %
% \author{
% \alignauthor
% Eric Berberich \\
%        \affaddr{Max-Planck Institute for Informatics}\\
%        \affaddr{Saarbr\"ucken, Germany}\\
%        \email{eric@mpi-inf.mpg.de}
% \alignauthor
% Pavel Emeliyanenko \\
%        \affaddr{Max-Planck Institute for Informatics}\\
%        \affaddr{Saarbr\"ucken, Germany}\\
%        \email{asm@mpi-inf.mpg.de}
% \alignauthor
% Michael Sagraloff \\
%        \affaddr{Max-Planck Institute for Informatics}\\
%        \affaddr{Saarbr\"ucken, Germany}\\
%        \email{msagralo@mpi-inf.mpg.de}
% }

\author{Eric Berberich%
\and Pavel Emeliyanenko
\and Michael Sagraloff}

\date{}
\maketitle

\vspace{0cm}
\begin{center}
\mails\\
  Max-Planck-Institut f\"ur Informatik, Saarbr\"ucken, Germany\\
\end{center}

%
%\authorrunning{E.~Berberich, P.~Emeliyanenko, M.~Sagraloff}
% (feature abused for this document to repeat the title also on left hand
% pages)

% the affiliations are given next; don't give your e-mail address
% unless you accept that it will be published
%
% NB: a more complex sample for affiliations and the mapping to the
% corresponding authors can be found in the file "llncs.dem"
% (search for the string "\mainmatter" where a contribution starts).
% "llncs.dem" accompanies the document class "llncs.cls".
%

% \toctitle{Lecture Notes in Computer Science}
% \tocauthor{Authors' Instructions}
% \maketitle
\ignore{
\eric{Bivariate Real Root Solving - efficient even with projection}

\eric{Eliminating typical drawbacks of bivariate elimination methods}

\eric{How to eliminate the typical drawbacks of elimination methods for bivariate system solving}

\eric{Eliminating one variable - without drawbacks}

\eric{Eliminating one variable -Eliminating drawbacks}

\eric{Eliminating drawbacks when eliminating one variable}
}

\begin{abstract}
We present an exact and complete algorithm to isolate the real solutions
of a zero-dimensional bivariate polynomial system. The proposed algorithm 
constitutes an elimination method which improves upon existing approaches 
in a number of points. First, the amount of purely symbolic operations is 
significantly reduced, that is, only resultant computation and square-free 
factorization is still needed. Second, our algorithm neither assumes 
generic position of the input system nor demands for any change of the 
coordinate system. The latter is due to a novel inclusion predicate to 
certify that a certain region is isolating for a solution.
Our implementation exploits graphics hardware to expedite the resultant 
computation. Furthermore,
we integrate a number of filtering techniques to improve the 
overall performance. Efficiency of the proposed method is proven by a 
comparison of our implementation with two state-of-the-art
implementations, that is, \lgp and Maple's \isolate.
For a series of challenging benchmark instances, experiments 
show that our implementation outperforms both contestants.
\end{abstract}

% \category{I.1.2}{Symbolic and Algebraic Manipulation}{Algorithms}[Algebraic algorithms]
% \terms{Algorithms, Experimentation, Measurement, Performance, Theory}
% \keywords{Bivariate Polynomial System, Real Roots, }

\section{Introduction}
\label{sec:introduction} 

Finding the real solutions of a bivariate polynomial system
is a fundamental problem with numerous applications in computational
geometry, computer graphics and computer aided geometric design. In
particular, topology and arrangement computations for algebraic
curves~\cite{eigenwilligkw07,eigenwilligk08,clpprt-topology,gn-efficient} crucially rely on the computation of common intersection
points of the given curves (and also the curves defined by their partial
derivatives). For the design of robust and certified
algorithms, we aim for exact methods to 
determine isolating regions for all solutions. Such methods should be
capable of handling any input, that is, even systems with multiple solutions.
The proposed algorithm \bs constitutes such an exact
and complete approach. Its input is a \emph{zero-dimensional} (i.e., there
exist only finitely many solutions) polynomial system $f(x,y)=g(x,y)=0$
defined by two bivariate polynomials with integer coefficients.
\bs computes disjoint boxes $B_1,\ldots,B_m\subset\RR^2$ for
\emph{all real
solutions}, where each box $B_i$ contains exactly one
solution (i.e., $B_i$ is isolating). In addition, the boxes can be refined
to an arbitrary small size.\\

\noindent\textbf{Main results.} \bs constitutes a classical 
elimination method which follows the same basic idea as the GRID method from~\cite{det-asymptotic} and the hybrid method proposed in~\cite{hh-xydirection}. More precisely, in a first step, the variables $x$ and $y$ are separately eliminated by
means of a resultant computation. Then, in the second step, for each possible candidate (represented as pair of projected solutions in $x$- and $y$-direction), we check whether it actually constitutes a solution of the given system or not. The
proposed method comes with a number of improvements compared to the aforementioned approaches and also to other existing elimination techniques~\cite{abrw-zeros,khfta-solvingsystems,r-rur,eigenwilligk08}. 
First, we tremendously reduced the amount of purely symbolic computations, 
namely, our method only demands for resultant computation and square-free 
factorization of univariate polynomials with integer coefficients.
Second, our implementation profits from a novel
approach~\cite{emel-ica3pp-10, emel-pasco-10} to compute
resultants exploiting the power of Graphics Processing Unite (GPUs). We
remark that, in comparison to the classical resultant computation on the
CPU, the GPU implementation is typically more than $100$-times faster. Our
experiments show that, for the considered instances, the resultant 
computation is no longer a ``global'' bottleneck of an elimination
approach.
Third, the proposed method never uses any kind of a coordinate 
transformation, even for non-generic input.\footnote{The 
system $f=g=0$ is non-generic if there exist two solutions sharing a 
common coordinate.} The latter is due to a novel inclusion predicate which 
combines information from the resultant computation and a homotopy 
argument to prove that a certain candidate box is isolating for a solution. 
Since we never apply a change of coordinates, our method
particularly profits in the case where $f$ and
$g$ are sparse or where we are only interested in ``local'' solutions
within a given box. Finally, we integrated a series of additional
filtering techniques which allow us to significantly speed up the computation 
for many instances. 

We implemented our algorithm as a prototypical package of 
\textsc{Cgal}~\cite{cgal-3.6} and ran our software on numerous challenging benchmark instances. 
For comparison, we considered two currently state-of-the-art 
implementations, that is, \isolate (based on \rs by Fabrice 
Rouillier with ideas from~\cite{r-rur}) from Maple~13 and 
\lgp by Xiao-Shan~Gao~\etal~\cite{LGP-09}.
Our experiments show that our method is efficient as it 
outperforms both contestants for most instances. More 
precisely, our method is comparable for all considered instances and
typically between $5$ and $10$-times faster. For some instances, 
we even improve by a factor of $50$ and more. Our filters apply to many
input systems and crucially contribute to the overall performance. 
We further remark that the gain in performance is not solely due to the 
resultant computation on the GPU but rather due to the combination of the 
sparse use of purely symbolic computations and efficient (approximate) 
subroutines. We prove the latter fact by providing 
running times with and without fast GPU-resultant computation.\\

\noindent\textbf{Related Work.} 
Since polynomial root solving is such an important problem in several
fields, plenty of distinct approaches exist and many textbooks are
dedicated to this subject. In general, we distinguish between two kinds 
of methods.

The first comprises non-certified or non-complete methods which 
give, in contrast to our goal here,
no guarantee on correctness or termination (e.g., if multiple roots
exists). Representatives of this category are numerical
(e.g. homotopy methods~\cite{sw-numerical}) or subdivision 
methods\footnote{Subdivision methods can be made certifying and 
complete when considering worst case separation bounds for the solutions, 
an approach which has not shown effective in practice so far.} 
(e.g.,~\cite{mp-subdivision,bcgy-complete,mourrain-sub}). 
A major strength of these methods is that
they are very efficient for most instances due to their use of approximate
computations such as provided by IntBis, ALIAS, IntLab or MPFI. 

The second category consists of certified and complete methods,
to which ours is to be added. So far,
only elimination methods based on \emph{(sparse) resultants}, 
\emph{rational univariate representation}, \emph{Groebner bases} 
or \emph{eigenvalues} have proven to be efficient representatives of 
this category; see, for instance, \cite{petit-alggeo,clo-using,
sturmfels-solving, yap-fundamental} for introductions to such symbolic
approaches. Common to all these methods is
that they combine a projection and a lifting step similar to the proposed 
approach. Recent exact and
complete implementations for computing the topology of algebraic curves and
surfaces~\cite{eigenwilligkw07,gn-efficient,bks-efficient} also make use of such elimination
techniques. However, already this low dimensional
application shows the main drawback of elimination methods, that is, they
tremendously suffer from costly symbolic computations. Furthermore, the given 
system might be in non-generic position which makes
the lifting step non-trivial. In such ``hard situations'', the
existing approaches perform a coordinate transformation (or project in
generic direction) which eventually increases the complexity of the input
polynomials. In particular, if we are only interested in ``local''
solutions within a given box, such methods induce a huge overhead of
purely symbolic computations. The proposed algorithm constitutes a 
contribution in two respects: The number of 
symbolic steps are crucially reduced and partially (resultant computation) 
outsourced to the GPU. In addition, generic and non-generic situations are 
treated in the same manner and, thus, a coordinate transformation which 
induces an overhead of symbolic computations is no longer needed.

\section{Setting}
\label{sec:setting}
The \emph{input} of our algorithm is the following polynomial system
\begin{align}
f(x,y)=\sum_{i,j\in\N:i+j\le m} f_{ij}x^iy^j=0,\quad
g(x,y)=\sum_{i,j\in\N:i+j\le n} g_{ij}x^iy^j=0,\label{system}
\end{align}
where $f$, $g\in\Z[x,y]$ are polynomials of total degrees $m$ and $n$, 
respectively. We also write\vspace{-0.1cm}
$$
f(x,y)=\sum_{i=0}^{m_x} f_{i}^{(x)}(y) x^i=\sum_{i=0}^{m_y} f_{i}^{(y)}(x)
y^i
\text{\hspace{0.1cm} and\hspace{0.1cm} } 
g(x,y)=\sum_{i=0}^{n_x} g_{i}^{(x)}(y)x^i=\sum_{i=0}^{n_y}
g_{i}^{(y)}(x)y^i,\vspace{-0.1cm}
$$
where $f_{i}^{(y)}$, $g_{i}^{(y)}\in\Z[x]$, $f_{i}^{(x)}$, 
$g_{i}^{(x)}\in\Z[y]$ and $m_x$, $n_x$ and $m_y$, 
$n_y$ denote the degrees of $f$ and $g$ considered as polynomials in $x$
and
$y$, respectively. Throughout the paper, it is assumed that $f$ and 
$g$ have no common factors.\footnote{Otherwise, $f$ and $g$ have to be decomposed into common and non-common factors (not part of our algorithm).} Hence, the set
$V_{\C}:=\{(x,y)\in\C^2|f(x,y)=g(x,y)=0\}$ of
(complex) solutions of~(\ref{system})
is zero-dimensional and consists, by B\'{e}zout's theorem, of at most
$m\cdot n$ distinct elements.

Our algorithm \emph{outputs} disjoint boxes
$B_k\subset\RR^2$ such that the union of all $B_k$ contains all \emph{real}
solutions
$$
V_{\RR}:=\{(x,y)\in\RR^2|f(x,y)=g(x,y)=0\}=V_{\C}\cap\RR^{2}
$$
of (\ref{system}) and each $B_k$ is \emph{isolating}, that is, it contains
exactly one solution.\vspace{0.20cm}

\noindent\textbf{Notation.} For an interval 
$I=(a,b)\subset\RR$, $m_{I}:=(a+b)/2$ denotes the \emph{center} and 
$r_{I}:=(b-a)/2$ the \emph{radius} of $I$. For an arbitrary 
$m\in\C$ and $r\in\RR^{+}$, $\Delta_{r}(m)$ denotes the disc with 
center $m$ and radius $r$.

\ignore{
\eric{This is a comment by eric}

\michael{This is a comment by michael}

\pavel{This is a comment by pavel}

\lncs{This is a comment by lncs}
}

\section{The Algorithm}
\label{sec:algorithm}
 \subsection{Resultants}\label{sec:resultants}

Our algorithm is based on well known elimination techniques, namely,
to consider the projections\vspace{-0.1cm}
$$
V^{(x)}_{\C}:=\{x\in\C|\exists y\in\C\text{ with }f(x,y)=g(x,y)=0\},\quad
V^{(y)}_{\C}:=\{y\in\C|\exists x\in\C\text{ with }f(x,y)=g(x,y)=0\}\vspace{-0.1cm}
$$
of all complex solutions $V_{\C}$ onto the $x$- and $y$-coordinate.
Resultant computation is a well studied tool to obtain an algebraic
description of these projection sets, that is, polynomials whose roots are
exactly the projections of the solution set $V_{\C}$. The resultant
$R^{(y)}=\operatorname{res}(f,g,y)$ of $f$ and $g$ with respect to the
variable $y$ is the determinant of the $(m_y+n_y)\times(m_y+n_y)$
\emph{Sylvester matrix}:
\[\scriptsize
S^{(y)}(f,g):=\left[
\begin{array}{ccccccc}
f_{m_y}^{(y)} & f_{m_y-1}^{(y)} & \ldots & f_0^{(y)} & 0\ \ldots & 0 \\
\vdots & \ddots & \ddots & & \ddots &  \vdots \\
0 & \ldots\ \ 0 & f_{m_y}^{(y)} & f_{m_y-1}^{(y)} & \ldots & f_0^{(y)} \\
g_{n_y}^{(y)} & g_{n_y-1}^{(y)} & \ldots & g_0^{(y)} & 0\  \ldots & 0 \\
\vdots & \ddots & \ddots & & \ddots &  \vdots \\
0 & \ldots\ \ 0 &  g_{n_y}^{(y)} & g_{n_y-1}^{(y)} & \ldots & g_0^{(y)}
\end{array}\right]
\]\normalsize
\noindent From the definition, it follows that $R^{(y)}\in\Z[x]$ has degree
less than or equal to $m\cdot n$. The resultant
$R^{(x)}=\operatorname{res}(f,g,x)$ of $f$ and $g$ with respect to $x$ is
defined in completely analogous manner by considering $f$ and $g$ as
polynomials in $x$ instead of $y$.
As mentioned above the resultant polynomials have the following important
property (cf.~\cite{bpr-algorithms} for a proof):
\begin{theorem}\label{thm:resultants}
The roots of $R^{(y)}$ and $R^{(x)}$ are exactly the projections of the
solutions of (\ref{system}) onto the $x$- and $y$-coordinate, respectively.
More precisely, $
V^{(x)}_{\C}=\{x\in\C|R^{(y)}(x)=0\}$ and $V^{(y)}_{\C}=\{y\in\C|R^{(x)}(y)=0\}.
$
The multiplicity of a root $\alpha$ of $R^{(y)}$ ($R^{(x)}$) is the sum of
the intersection multiplicities\footnote{The multiplicity of a solution $(x_0,y_0)$ of ($\ref{system}$) is defined as the dimension of the localization of 
$\C[x,y]/(f,g)$ at $(x_0,y_0)$ considered as $\C$-vector space  (cf.\cite[p.148]{bpr-algorithms})} of all solutions of
(\ref{system}) with $x$-coordinate ($y$-coordinate) $\alpha$.
\end{theorem}

\subsection{Isolating the Solutions: $\textsc{Project}$, $\textsc{Separate}$ and $\textsc{Validate}$}\label{sec:algo}

We start with the following high level description of the proposed
algorithm which decomposes into three subroutines:
In the first step (\textsc{Project}), we project the complex solutions
$V_{\C}$ of (\ref{system})
onto the $x$- and onto the $y$-axis. More precisely, we compute the restrictions $
V^{(x)}_{\RR}:=V^{(x)}_{\C}\cap \RR$ and $V^{(y)}_{\RR}:=V^{(y)}_{\C}\cap \RR.
$
of the complex projection sets $V^{(x)}_{\C}$ and $V^{(y)}_{\C}$ to
the real axes and isolating intervals for their elements.
Obviously, the real solutions $V_{\RR}$ are contained in the cross product $\mathcal{C}:=V^{(x)}_{\RR}\times 
V^{(y)}_{\RR}\subset\RR^2$. 
In the second step (\textsc{Separate}), we compute isolating discs
which well separate the projected solutions from each other. The latter prepares the third step (\textsc{Validate}) in which 
candidates of $\mathcal{C}$ are either discarded or certified 
to be a solution of (\ref{system}). 
Our \emph{main theoretical contribution} is the introduction of a 
novel predicate to ensure that a certain candidate 
$(\alpha,\beta)\in\mathcal{C}\cap V_{\RR}$ actually fulfills 
$f(\alpha,\beta)=g(\alpha,\beta)=0$ (cf.~Theorem~\ref{thm:inclusion}).
For all candidates $(\alpha,\beta)\in\mathcal{C}\backslash V_{\RR}$,
simple interval arithmetic suffices to exclude $(\alpha,\beta)$ as
a solution of (\ref{system}).

We remark that,
in order to increase the efficiency of our implementation,
we also introduce additional filtering techniques to eliminate
many of the candidates in $\mathcal{C}$.
However, for the sake of clarity, we refrain from integrating our
filtering techniques in the following description
of the three subroutines. 
Filtering techniques are covered separately in Section~\ref{ssec:filters}.
Section~\ref{ssec:gpu_res} briefly discusses a
highly parallel algorithm on the graphics hardware
to accelerate computations of the resultants needed in the first step.\vspace{0.25cm}

\noindent\emph{\textsc{Project}:} We compute the resultant
$R:=R^{(y)}=\operatorname{res}(f,g,y)
\in\Z[x]$ and a square-free
factorization of~$R$. More precisely, we determine square-free and pairwise 
coprime factors $r_{i}\in\Z[x]$, $i=1,\ldots,\deg(R)$, such that
$R(x)=\prod_{i=1}^{\deg
(R)}\left(r_{i}(x)\right)^{i}$. We remark that, for some
$i\in\{1,\ldots,\deg(R)\}$, $r_{i}(x)=1$.
Yun's algorithm~\cite[Alg.~14.21]{gathen} constructs such a square-free factorization by
essentially computing greatest common divisors of $R$ and its higher
derivatives in an iterative way.
Next, we isolate the real roots $\alpha_{i,j}$, $j=1,\ldots,\ell_{i}$, of
the polynomials $r_{i}$. That is, we determine disjoint isolating intervals
$I(\alpha_{i,j})\subset\RR$ such that each interval
$I(\alpha_{i,j})$ contains exactly one root (namely, $\alpha_{i,j}$) of
$r_{i}$ and the union
of all $I(\alpha_{i,j})$, $j=1,\ldots,\ell_{i}$, covers all real roots of
$r_{i}$.
For the real root isolation, we consider the Descartes method~\cite{vca,RS} as a suited
algorithm.
From the square-free factorization we know that $\alpha_{i,j},
j=1,\ldots,\ell_{i}$, is a root of $R$ with multiplicity~$i$.\vspace{0.25cm}

\noindent\emph{\textsc{Separate}:} We separate the real roots
of $R=R^{(y)}$ from all other (complex) roots of $R$, a step which is
crucial for the final validation. More precisely, let
$\alpha=\alpha_{i_{0},j_{0}}$ be the $j_{0}$-th real root of the polynomial
$r_{i_{0}}$, where $i_{0}\in\{1,\ldots,\deg(R)\}$ and
$j_{0}\in\{1,\ldots,\ell_{i_{0}}\}$ are arbitrary indices. We refine the
corresponding isolating interval $I=(a,b):=I(\alpha)$ such that the disc
$\Delta_{8r_{I}}(m_I)$ does not contain any root of $R^{(y)}$
except~$\alpha$.
For the refinement of $I$, we use quadratic interval 
refinement~\cite{abbott-qir-06,KerberCASC09} (QIR) which constitutes a highly efficient 
method because of its simple tests and the fact that it eventually achieves
quadratic convergence.

In order to test whether the disc $\Delta_{8r_{I}}(m_I)$ isolates $\alpha$
from all other roots of $R$, we introduce a novel method based on the
following test:
\begin{equation*}
	T^p_K(m,r):|p(m)| - K \sum_{k\ge 1}
		\left|\frac{p^{(k)}(m)}{k!} \right|r^k>0,
\end{equation*}
where $p\in\RR[x]$ denotes an arbitrary polynomial and $m$, $r$, $K$
arbitrary real values. Then, the following theorem holds
(cf.~Appendix~\ref{thm:testappendix} for a proof):

\begin{theorem}
\label{thm:test}
Consider a disk $\Delta=\Delta_m(r)\subset\C$ with center $m$ and radius
$r$.
\begin{enumerate}
\item If $T_{K}^p(m,r)$ holds for some $K\geq 1$, then the closure
$\overline{\Delta}$ of $\Delta$ contains no root of $p$.
\item	If $T^{p'}_{K}(m,r)$ holds for a $K\geq \sqrt{2}$, then
$\overline{\Delta}$ contains at most one root of $p$.
\end{enumerate}
\end{theorem}

Theorem~\ref{thm:test} now directly applies to the above scenario. More
precisely, $I$ is refined until $T^{(r_{i_{0}})'}_{3/2}(m_I,8r_I)$ and
$T^{r_{i}}_1(m_I,8r_I)$ holds for all $i\neq i_0$.
If the latter is fulfilled, $\Delta_{8r_I}(m_I)$ isolates $\alpha$ from all
other roots of $R$. In this situation, we obtain a lower bound $LB(\alpha)$ for $|R(z)|$ on the boundary of
$\Delta(\alpha):=\Delta_{2r_{I}}(m_{I})$:

\begin{lemma}\label{lem:lowerbound}
The disc $\Delta(\alpha)=\Delta_{2r_{I}}(m_{I})$ isolates $\alpha$ from all
other (complex) roots of $R$ and, for any $z$ on the boundary $\partial
\Delta(\alpha)$ of $\Delta(\alpha)$, it holds that $|R(z)|>LB(\alpha):= 2^{-i_0-\deg(R)}|R(m_I-2r_I)|$.
\end{lemma}
 
\begin{proof}
$\Delta(\alpha)$ is isolating as already $\Delta_{8r_{I}}(m_{I})$ is
isolating. Then, let $\beta\neq \alpha$ be an arbitrary root of $R$ and
$d:=|\beta-m_I|>8r_I$ the distance between $\beta$ and $m_I$. Then, for any
point $z\in\partial\Delta(\alpha)$, it holds that
$$\frac{|z-\beta|}{|(m_I-2r_I)-\beta|}>\frac{d-2r_I}{d+2r_I}=1-\frac{4r_I}{
d+2r_I}>\frac{1}{2}\hspace{0.2cm}\text{ and }\hspace{0.2cm}
\frac{|z-\alpha|}{|(m_I-2r_I)-\alpha|}>\frac{r_I}{3r_I}>\frac{1}{4}.
$$
Hence, it follows that
$$\frac{|R(z)|}{|R(m_I-2r_I)|}>\prod_{\beta\neq\alpha: 
R(\beta)=0}\frac{|z-\beta|}{|(m_I-2r_I)-\beta|}\cdot 
\left(\frac{|z-\alpha|}{|(m_I-2r_I)-\alpha|}\right)^{i_{0}}>4^{-i_{0}}
2^{-\deg(R)+i_{0}},$$
where each root $\beta$ occurs as many times in the above product as its
multiplicity as a root of $R$.
\end{proof} 
 
We evaluate $LB(\alpha)=2^{-i_{0}-\deg(R)}|R(m_I-2r_I)|$
and store the interval $I(\alpha)$, the disc $\Delta(\alpha)$ and the lower
bound $LB(\alpha)$ for $|R(z)|$ on the boundary $\partial\Delta(\alpha)$ of
$\Delta(\alpha)$.\\

Proceeding in exactly the same manner for each real
root $\alpha$ of $R^{(y)}$, we get an isolating interval $I(\alpha)$,
an isolating disc $\Delta(\alpha)=\Delta_{2r_{I}}(m_{I})$ and a lower bound
$LB(\alpha)$ for $|R^{(y)}|$ on $\partial\Delta(\alpha)$.
For the resultant polynomial $R^{(x)}$, \textsc{Project} and
\textsc{Separate} are processed in exactly the same manner: We compute
$R^{(x)}$ and a corresponding square-free factorization. Then, for each
real root $\beta$ of $R^{(x)}$, we compute a corresponding isolating
interval $I(\beta)$, a disc $\Delta(\beta)$ and a lower bound $LB(\beta)$
for $|R^{(x)}|$ on $\partial\Delta(\beta)$.\vspace{0.25cm}

\noindent\emph{\textsc{Validate}:}  We start with the following theorem:

\begin{theorem}\label{thm:boxproperties}
Let $\alpha$ and $\beta$ be arbitrary real roots of $R^{(y)}$ and
$R^{(x)}$, respectively. Then,
\begin{enumerate}
\item the polydisc $\Delta(\alpha,\beta):=\Delta(\alpha)
\times\Delta(\beta)\subset\C^2$
\ignore{and, thus, the box $$B(\alpha,\beta):=I(\alpha)\times
I(\beta)\subset\RR^2$$}contains at most one (complex) solution of
(\ref{system}). If $\Delta(\alpha,\beta)$ contains a 
solution of (\ref{system}), then this solution is real valued and equals
$(\alpha,\beta)$.
\item For an arbitrary point $(z_1,z_2)\in\C^2$ on the boundary of
$\Delta(\alpha,\beta)$, it holds that
\begin{align*}
&|R^{(y)}(z_1)|>LB(\alpha)\text{ if }z_1\in\partial\Delta(\alpha)\text{, and }
|R^{(x)}(z_2)|>LB(\beta) \text{ if }z_2\in\partial\Delta(\beta).
\end{align*}
\end{enumerate}
\end{theorem}

\begin{proof}
(1) is an easy consequence from the construction of the discs
$\Delta(\alpha)$ and $\Delta(\beta)$. Namely, if $\Delta(\alpha,\beta)$
contains two distinct solutions of (\ref{system}), then they would differ
in at least one variable. Thus, one of the discs $\Delta(\alpha)$ or
$\Delta(\beta)$ would contain two roots of $R^{(y)}$ or $R^{(x)}$. Since
both discs are isolating for a root of the corresponding resultant
polynomial, it follows that $\Delta(\alpha,\beta)$ contains at most one
solution. In the case, where $\Delta(\alpha,\beta)$ contains a solution of
(\ref{system}), this solution must be real since, otherwise,
$\Delta(\alpha,\beta)$ would also contain a corresponding complex conjugate
solution ($f$ and $g$ have real valued coefficients).
(2) follows directly from the definition of $\Delta(\alpha,\beta)$, the
definition of $LB(\alpha)$, $LB(\beta)$ and Lemma~\ref{lem:lowerbound}.
\end{proof}

We denote $B(\alpha,\beta)=I(\alpha)\times I(\beta)\subset\RR^2$ a
\emph {candidate box} for a real solution of (\ref{system}), where $\alpha$
and $\beta$ are real roots of $R^{(y)}$ and $R^{(x)}$, respectively. Due to
Theorem~\ref{thm:boxproperties}, the corresponding "container polydisc"
$\Delta(\alpha,\beta)\subset \C^2$ either contains no solution of
(\ref{system}) or $(\alpha,\beta)$ is the only solution contained in
$\Delta(\alpha,\beta)$. Hence, for each candidate pair
$(\alpha,\beta)\in\mathcal{C}$, it suffices to show that either
$(\alpha,\beta)$ is no solution of (\ref{system}) or the corresponding
polydisc $\Delta(\alpha,\beta)$ contains at least one solution.
In the following steps, we fix the polydiscs $\Delta(\alpha,\beta)$ whereas
the boxes $B(\alpha,\beta)$ are further refined (by further refining the
isolating intervals $I(\alpha)$ and $I(\beta)$). We also introduce
exclusion and inclusion predicates such that, for sufficiently small
$B(\alpha,\beta)$, either $(\alpha,\beta)$ can be discarded or certified as
a solution of (\ref{system}).\\

\label{pref:interval_exclusion}
In order to \emph{exclude} a candidate box, we use simple interval arithmetic.
More precisely, we evaluate $\Box f(B(\alpha,\beta))$ and $\Box
g(B(\alpha,\beta))$, where $\Box f$ and $\Box g$ constitute box functions
for $f$ and $g$, respectively: If either $\Box
f(B(\alpha,\beta))$ or $\Box g(B(\alpha,\beta))$ does not contain zero,
then $(\alpha,\beta)$ cannot be a solution of (\ref{system}). Vice versa,
if $(\alpha,\beta)$ is not a solution and $B(\alpha,\beta)$ becomes
sufficiently small, then either $0\notin\Box f(B(\alpha,\beta))$ or
$0\notin\Box g(B(\alpha,\beta))$ and our exclusion predicate applies.\vspace{0.20cm}

It remains to provide an \emph{inclusion predicate}, that is, a method to
ensure that a certain candidate $(\alpha,\beta)\in\mathcal{C}$ is
actually a solution of (\ref{system}). We first rewrite the resultant polynomial $R^{(y)}$ as\vspace{-0.1cm}
\begin{align*}
R^{(y)}(x)=u^{(y)}(x,y)\cdot f(x,y)+v^{(y)}(x,y)\cdot g(x,y),\vspace{-0.1cm}
\end{align*}
where $u^{(y)}$, $v^{(y)}\in\Z[x,y]$. Furthermore, $u^{(y)}$ and $v^{(y)}$ can be
expressed as determinants of "Sylvester-like" matrices
$U^{(y)}$ and $V^{(y)}$.
\ignore{:
\[\small
U^{(y)}=\left|
\begin{array}{ccccccc}
f_{m_y}^{(y)} & f_{m_y-1,y}^{(y)} & \ldots & f_{0}^{(y)} & 0\ \ldots &
y^{n_y-1} \\
\vdots & \ddots & \ddots & & \ddots &  \vdots \\
0 & \ldots\ \ 0 & f_{m_y}^{(y)} & f_{m_y-1}^{(y)} & \ldots & 1 \\
g_{n_y}^{(y)} & g_{n_y-1}^{(y)} & \ldots & g_{0}^{(y)} & 0\  \ldots & 0 \\
\vdots & \ddots & \ddots & & \ddots &  \vdots \\
0 & \ldots\ \ 0 &  g_{n_y}^{(y)} & g_{n_y-1}^{(y)} & \ldots & 0
\end{array}\right|,\hspace{0.2cm}
V^{(y)} =\left|
\begin{array}{ccccccc}
f_{m_y}^{(y)} & f_{m_y-1}^{(y)} & \ldots & f_{0}^{(y)} & 0\ \ldots & 0 \\
\vdots & \ddots & \ddots & & \ddots &  \vdots \\
0 & \ldots\ \ 0 & f_{m_y}^{(y)} & f_{m_y-1}^{(y)} & \ldots & 0 \\
g_{n_y}^{(y)} & g_{n_y-1}^{(y)} & \ldots & g_{0}^{(y)} & 0\  \ldots &
y^{m_y-1} \\
\vdots & \ddots & \ddots & & \ddots &  \vdots \\
0 & \ldots\ \ 0 &  g_{n_y}^{(y)} & g_{n_y-1}^{(y)} & \ldots & 1
\end{array}\right|.
\]}
More precisely, $U^{(y)}$ and
$V^{(y)}$ are obtained from $S^{(y)}(f,g)$ by replacing the last
column with vectors $(y^{n_y-1}\dots 1\ 0 \dots 0)^T$ and $(0 \dots 0\
y^{m_y-1}\dots 1)^T$ of appropriate size, respectively~\cite[p.287]{algs-92}. 
Both matrices have size
$(n_y+m_y)\times(n_y+m_y)$ and univariate polynomials in
$x$ (the first $n_{y}+m_{y}-1$ columns) or powers of $y$ (only the last
column) or zeros as entries. We now aim for upper bounds for $|u^{(y)}|$ and
$|v^{(y)}|$ on the polydisc $\Delta(\alpha,\beta)$. The polynomials $u^{(y)}$ 
and $v^{(y)}$ have huge coefficients and their computation, either via a 
signed remainder sequence or via determinant evaluation, is very costly. 
Hence, we directly derive 
such upper bounds from the corresponding matrix representations 
\textbf{without computing} $u^{(y)}$ and
$v^{(y)}$: Due to 
Hadamard's
bound, $|u^{(y)}|$ is smaller than the product of the $2$-norms of
the column vectors of $U^{(y)}$. The absolute value of each of the entries
of $U^{(y)}$ can be easily upper bounded by using interval arithmetic on a
box in $\C^2$ that contains the polydisc $\Delta(\alpha,\beta)$. Hence, we get
an upper bound on the $2-$norm of each column vector and, thus, an upper
bound $UB(\alpha,\beta,u^{(y)})$ for $|u^{(y)}|$ on $\Delta(\alpha,\beta)$
by multiplying the bounds for the column vectors. In the same manner, we also
derive an upper bound $UB(\alpha,\beta,v^{(y)})$ for $|v^{(y)}|$ on
$\Delta(\alpha,\beta)$. With respect to our second projection direction, we
write $R^{(x)}=u^{(x)}\cdot f+v^{(x)}\cdot g$
with corresponding polynomials $u^{(x)}$, $v^{(x)}\in\Z[x,y]$. In exactly
the same manner as done for $R^{(y)}$, we compute corresponding upper
bounds $UB(\alpha,\beta,u^{(x)})$ and $UB(\alpha,\beta,v^{(x)})$ for
$|u^{(x)}|$ and $|v^{(x)}|$ on $\Delta(\alpha,\beta)$. 

\begin{theorem}\label{thm:inclusion}
If there exists an $(x_0,y_0)\in\Delta(\alpha,\beta)$ with
\begin{align}
&UB(\alpha,\beta,u^{(y)})\cdot |f(x_0,y_0)|+UB(\alpha,\beta,v^{(y)})\cdot
|g(x_0,y_0)|< LB(\alpha) \text{ and }\label{ineq1}\\
&UB(\alpha,\beta,u^{(x)})\cdot |f(x_0,y_0)|+UB(\alpha,\beta,v^{(x)})\cdot
|g(x_0,y_0)|< LB(\beta)\label{ineq2},
\end{align}
then $\Delta(\alpha,\beta)$ contains a solution of (\ref{system}) and,
thus, $f(\alpha,\beta)=0$.
\end{theorem}

\begin{proof}
The proof uses a homotopy argument. Namely, we consider the parameterized
system
\begin{align}
f(x,y)-(1-t)\cdot f(x_0,y_0)=g(x,y)-(1-t)\cdot g(x_0,y_0)=0,\label{systemt}
\end{align}
where $t$ is an arbitrary real value in $[0,1]$. For $t=1$, (\ref{systemt})
is equivalent to our initial system (\ref{system}). For $t=0$,
(\ref{systemt}) has a solution in $\Delta(\alpha,\beta)$, namely,
$(x_0,y_0)$. The complex solutions of (\ref{systemt}) continuously depend
on the parameter $t$. Hence, there exists a ``solution path''
$\Gamma:[0,1]\mapsto\C^2$ which connects $\Gamma(0)=(x_0,y_0)$ with a
solution $\Gamma(1)\in\C^2$ of (\ref{system}). We show that $\Gamma(t)$
does not leave the polydisc $\Delta(\alpha,\beta)$ and, thus,
(\ref{system}) has a solution in $\Delta(\alpha,\beta)$: Assume that the path
$\Gamma(t)$ leaves the polydisc, then there exists a $t'\in[0,1]$ with
$(x',y')=\Gamma(t')\in\partial\Delta(\alpha,\beta)$. We assume that
$x'\in\partial\Delta(\alpha)$ (the case $y'\in\partial\Delta(\beta)$ is
treated in analogous manner). Since $(x',y')$ is a solution of
(\ref{systemt}) for $t=t'$, we must have $|f(x',y')|\le|f(x_0,y_0)|$ and
$|g(x',y')|\le|g(x_0,y_0)|$. Hence, it follows that
\begin{align*}
|R^{(y)}(x')|&=|u^{(y)}(x',y') f(x',y')+v^{(y)}(x',y')
g(x',y')|
\le|u^{(y)}(x',y')|\cdot |f(x',y')|+|v^{(y)}(x',y')|\cdot |g(x',y')|\\
&\le UB(\alpha,\beta,u^{(y)})\cdot
|f(x_0,y_0)|+UB(\alpha,\beta,v^{(y)})\cdot |g(x_0,y_0)|<LB(\alpha).
\end{align*}
This contradicts the fact that $|R^{(y)}(x')|$ is lower bounded by
$LB(\alpha)$. It follows that $\Delta(\alpha,\beta)$ contains a solution of
(\ref{system}) and, according to Theorem~\ref{thm:boxproperties}, this
solution must be $(\alpha,\beta)$.
\end{proof}

Theorem~\ref{thm:inclusion} now directly applies as an inclusion predicate.
Namely, in each refinement of $B(\alpha,\beta)$, we choose an arbitrary
$(x_0,y_0)\in B(\alpha,\beta)$ (e.g., the center
$(m_{I(\alpha)},m_{I(\beta)})$) of the candidate box $B(\alpha,\beta)$) and
check whether both inequalities (\ref{ineq1}) and (\ref{ineq2}) are
fulfilled. If $(\alpha,\beta)$ is a solution of (\ref{system}), then both
inequalities eventually hold and, thus, we have shown that $(\alpha,\beta)$
is a solution.\\

We remark that the upper bounds $UB(\alpha,\beta,u^{(y)})$, 
$UB(\alpha,\beta,v^{(y)})$, $UB(\alpha,\beta,u^{(x)})$ and 
$UB(\alpha,\beta,v^{(y)})$ are far from being optimal. Nevertheless, our 
inclusion predicate is still efficient since we can approximate the potential 
solution $(\alpha,\beta)$ with quadratic convergence due to QIR. Hence, the 
values $f(x_0,y_0)$ and $g(x_0,y_0)$ become very small after a few iterations. 
In order to improve the above bounds, we propose to consider more 
sophisticated methods from numerical analysis and matrix perturbation 
theory~\cite{IR-perturbation,rump-verifiedbounds}.
Finally, we would like to emphasize that our method applies particularly well 
to the 
situation where we are only interested in the solutions of (\ref{system}) 
within a given box
$\mathcal{B}=[A,B]\times[C,D]\subset\RR^{2}$. Namely, in \textsc{Project}, we 
only have to search for roots within the interval $[A,B]$ ($[C,D]$) for 
$R^{(y)}$ ($R^{(x)}$) and only candidate boxes within $\mathcal{B}$ have to be 
considered in \textsc{Seperate} and \textsc{Validate}.

\section{Speedups}
\label{sec:speedups}

\subsection{Resultants on graphics hardware}
\label{ssec:gpu_res}

Computing the resultants of bivariate polynomials is an important 
``symbolic part'' of our algorithm. Despite a large body of 
research existing on this subject, symbolic computations 
still constitute a large bottleneck in many algorithms and 
substantially limit their range of applicability. 
We use a novel approach exploiting the power of GPUs to 
dramatically reduce the time for computing resultants.
In this section, we briefly discuss the algorithm; we refer 
the reader to~\cite{emel-ica3pp-10, emel-pasco-10} for details.

\ignore{Needless to say that Graphics Processor Units (GPUs) currently 
offer the best parallel performance per processing unit cost ratio.}

Our approach is based on the classical ``divide-conquer-combine'' modular
algorithm by Collins~\cite{collins-71}.
The algorithm can be summarized in the following steps.
\begin{inparaenum}[1.]\item Apply modular and evaluation homomorphisms to
map the problem to computing a large set of problems over a simple domain.
\item Compute a set of univariate resultants over a prime field. \item
Recover the resultant through polynomial interpolation and Chinese
remaindering.\end{inparaenum}

Unfortunately, Collins' algorithm in its original form is not suitable for
a realization on the GPU. This is because the amount of parallelism exposed
by the modular approach is far too low to satisfy the needs of
massively-threaded architectures.
We deal with this issue by reducing the problem to computations with
\emph{structured matrices} because matrix operations typically 
map very well to the GPU's threading model.
As a result, all steps of the algorithm except the initial modular reduction and
partly the Chinese remaindering are run on the graphics hardware, thereby
minimizing the amount of work to be done on the CPU. For expository
purposes, we outline here the computation of univariate resultants in
more detail.

Suppose, $f$ and $g$ are polynomials in $\Z[x]$ of degrees $m$ and $n$
respectively.
It is clear that the resultant of $f$ and $g$ reduces to the triangular
factorization of the Sylvester matrix~$S$
(see Section~\ref{sec:resultants}). The matrix $S\in\Z^{r\times r}$ ($r =
m+n$) is structured as it satisfies the displacement
equation~\cite{displ-95}:
\[
S - Z_r S A^T = GB^T\mbox{, with }A=Z_m\oplus Z_n\mbox{ and }G, B\in\Z^{r\times 2}\mbox{,}
\]
\noindent here $Z_s\in\Z^{s\times s}$ is a down-shift matrix zeroed everywhere
except for 1's on the first subdiagonal. Accordingly, the \emph{generators} $G, B$
are matrices whose entries can be deduced from the matrix $S$ by
inspection. Hence, we can apply the generalized Schur algorithm 
which operates on the matrix generators to compute the matrix factorization
in $\mathcal{O}(r^2)$ time, see~\cite[p.~323]{displ-95}.

In short, the Schur algorithm is an iterative procedure: In each step, it
brings the matrix generators to a ``special form'' from which triangular
factors can easily be deduced based on the displacement equation. Using
division-free modifications this procedure can be efficiently performed in
a finite field giving rise to the factorization algorithm running in
$\mathcal{O}(r)$ time using $r$ processors.

Suppose that we have evaluated the polynomials $f,g\in\Z[x,y]$ as defined in (\ref{system}) at a number of points $x_i\in\Z_p$
and computed a set of univariate resultants over a prime field $\Z_p$, that is,
$z_i^{(p)} = \operatorname{res}(f(x_i,y),g(x_i,y),y)\in\Z_p$. Then, the resultant polynomial $R^{(y)}(x)$ is interpolated over the prime field $\Z_p$ and eventually lifted to an integer solution via Chinese remaindering. We remark that polynomial interpolation corresponds to solving the Vandermonde
system.\footnote{Here we are not concerned with the fact that Vandemonde
systems are notoriously ill-conditioned since all operations are performed
in a finite field.}
Again, exploiting the structure of Vandermonde matrix we can use the Schur
algorithm to solve the system in a small parallel time.

% \eric{some details to highlight increase of efficiency like ``On average it
% is 1-3 orders of magnitude faster''}\michael{it makes the overall algo 1-3
% times faster; but, more interesting is the fact that it makes the resultant
% computation 100-times faster; hence the resultant computation is not a
% bootleneck anymore; at least for the considered instances.}

% In our implementation we use an efficient modular arithmetic developed
% reflecting the architectural features of the GPU

\subsection{Filters}
\label{ssec:filters}

Besides the parallel resultant computation,
our algorithm elaborates some filtering techniques to 
early validate a majority of the candidates.

As first step, we group candidates along the same vertical line 
(a \emph{fiber})
at an $x$-coordinate~$\alpha$ (a root of $R^{(y)}$) to process them
together. This allows us to use extra information on the real roots 
of $f(\alpha,y)\in\RR[y]$ and $g(\alpha,y) \in \RR[y]$ for candidate
validation.
\begin{figure}[tb]
\centering\footnotesize\sffamily
\resizebox{!}{!}{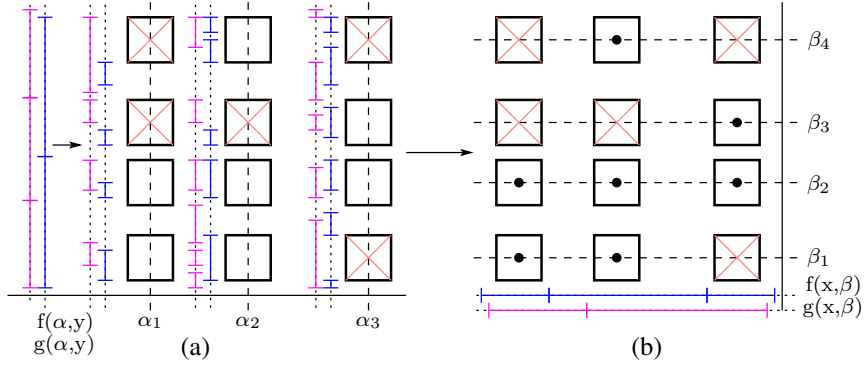} 
\caption{\textbf{(a)} Intervals containing the roots of $f(\alpha,y)$ and 
$g(\alpha,y)$ are refined until they either do not overlap or are fully 
included in candidate boxes. In the former case, the boxes can be 
discarded. \textbf{(b)} Unvalidated candidates are passed to 
\emph{bidirectional} filter
which runs bitstream isolation in another direcion}
\label{fig:BS}
\end{figure}
We replace the tests based on interval evaluation 
(see page~\pageref{pref:interval_exclusion}) by 
a test based on the \emph{bitstream Descartes}  
isolator~\cite{ek+-descartes} (\bdesc for short). 
This method allows us to isolate the real roots of a polynomial with 
``bitstream'' coefficients, that is, coefficients that can be approximated
to arbitrary precision. \bdesc starts from an interval guaranteed to 
contain all real roots of a polynomial, and proceeds with interval 
subdivisions giving rise to a \emph{subdivision tree}. 
Accordingly, the approximation precision for coefficients 
is increased in each step of the algorithm.
Each leaf of the tree is associated with an interval and stores an upper and a lower
bound on the number of real roots within this interval based on Descartes' Rule of Signs.
An interval is not further subdivided when both bounds equal $0$, 
where the interval is discarded, or $1$, where we have found an \emph{isolating interval}.
Isolating intervals can be refined to arbitrary precision. We remark that \bdesc terminates if all real roots are simple. Otherwise, intervals which contain a multiple root are further refined but never certified to contain a root.

In our algorithm, we apply \bdesc to the polynomials
$f(\alpha,y)$ and $g(\alpha,y)$. Eventually, intervals that do not 
share a common root of both polynomials will be discarded.
This property is essential for our ``filtered'' algorithm:
a candidate box $B(\alpha,\beta)$ can be rejected as soon as
the associated $y$-interval $I(\beta)$ does not overlap with \emph{at least} 
one of the isolating intervals associated with $f(\alpha,y)$ or 
$g(\alpha,y)$; see~Figure~\ref{fig:BS}~(a).

Grouping candidates along a fiber $x=\alpha$ also enables us to
use \emph{combinatorial} tests to discard or to certify them.
First, when the number of certified solutions reaches
$\mathrm{mult}(\alpha)$,
the remaining candidates are automatically discarded because 
each real solution contributes at least once to $\alpha$'s multiplicity 
as a root of $R^{(y)}$ (cf. Theorem~\ref{thm:resultants}). Second, if
$\mathrm{mult}(\alpha)$ is odd and all except one candidate along the fiber
are discarded, then the remaining candidate must be a real solution.
This is because complex roots come in conjugate pairs and, thus, 
do not change the parity of $\mathrm{mult}(\alpha)$.
We remark that, in case where the system (\ref{system}) is in
\emph{generic position} and the multiplicities of all roots of $R$ are odd,
the combinatorial test already suffices to certify all solutions 
without the need to apply our inclusion predicate.

Now, suppose that, after the combinatorial test, there are several 
candidates left along a fiber. For instance, the latter can indicate 
the presence of \emph{covertical} solutions. 
In this case, before using the inclusion predicate, 
we can apply the aforementioned filters in \emph{horizontal} direction.
More precisely, we construct the lists of unvalidated
candidates sharing the same $y$-coordinate $\beta$ and process them 
along a horizontal fiber. 
For this step, we initialize the bitstream trees for $f(x,\beta)$ and
$g(x,\beta)$ and proceed in exactly the same way as done for vertical 
fibers; see~Figure~\ref{fig:BS}~(b).
Candidates that still remain undecided after all tests are processed by
considering our inclusion predicate.
In Section~\ref{sec:experiments}, where we next examine the efficiency 
of our filters, we will refer to this procedure as the 
\emph{bidirectional} filter.

\section{Implementation \& Experiments}
\label{sec:experiments}
We have implemented our algorithm as a prototypical package of
\cgal.\footnote{Computational Geometry Algorithms Library,
\url{www.cgal.org}.}
As throughout the library we follow a \emph{generic programming paradigm}
that, for instance, enables us to easily exchange the number types used or
the method to isolate the roots of a polynomial without altering the
main structure of the implementation.
\begin{table}[t]
\centering\footnotesize\sffamily
\caption{Description of the curves used in the first part of experiments. 
In case only a single curve given, the second curve is taken to be the 
first derivative w.r.t.\ $y$-variable.}
\label{tbl:test-set1}
\begin{minipage}[t]{\linewidth}
\centering\footnotesize\sffamily
 \renewcommand{\thefootnote}{*}  
\begin{tabular}[]{|l|p{2.5in}||l|p{2.5in}|}
\hline
Instance  & Description & Instance  & Description \\[3pt]
\hline\hline
L4\_circles & circles w.r.t.\ L4-norm, clustered solutions & SA\_4\_4\_eps\footnotemark  & singular points with high tangencies, displaced  \\
\hline
curve\_issac & a curve appeared in~\cite{LGP-09} & FTT\_5\_4\_4\footnotemark & many non-rational singularities \\
\hline 
tryme & covertical solutions, many candidates to check & dfold\_10\_6\footnotemark & a curve with many half-branches  \\
\hline
large\_curves & large number of solutions & cov\_sol\_20 & covertical solutions  \\
\hline
degree\_6\_surf  & silhouette of an algebraic surface, covertical solutions in both directions  & mignote\_xy & a product of $x$/$y$- Mignotte polynomials, displaced; many clustered solutions\\ 
\hline
challenge\_12\footnotemark  & many candidate solutions to be checked & spider & degenerate curve, many clustered solutions  \\
\hline
\end{tabular}
\footnotetext{* These curves were taken from~\cite{labs_10}}
\addtocounter{footnote}{-4}
\end{minipage}
\end{table}
\begin{table}[ht]
\centering\footnotesize\sffamily
\caption{Experiments for the curves listed in Table~\ref{tbl:test-set1}.
Execution times are in seconds, including resultant computations. \bs-GPU: our 
approach with GPU-resultants; \bs-CPU: our approach with \cgal's 
CPU-resultants; \isolate and \lgp use Maple's implementation for the resultant computation.
Bold face indicates default setup for \bs.}
\label{tbl:test1}
\begin{minipage}[t]{\linewidth}
\centering
\begin{tabular}[]{|l| r | r||r | r|| r|r |r||r|r|} %p{2cm}|p{2cm}|p{2cm}|p{2cm}|p{2cm}|p{2cm}|}
\hline
 & & & \multicolumn{2}{|c||}{\textsc{BS+allfilters}} & \textsc{BS+bstr+comb} & \textsc{BS+bstr} & \textsc{BS} & \isolate & \lgp \\
Instance &$y$-degree & \#sols & CPU & GPU & \multicolumn{3}{c||}{GPU} & Maple & Maple \\
\hline
\hline
L4\_circles     & 16      &  17 &   2.74 &  \textbf{1.68}            &  1.52                       &   1.71 &   0.68   &    1.20 &      7.40 \\
\hline
curve\_issac    & 15      &  18 &   4.30 &  \textbf{3.21}            &  2.70                       &   3.47 &   1.84   &   70.91 &      3.54 \\
\hline
tryme           & 24, 34  &  20 &  98.56 & \textbf{29.31}            & 31.63                       &  89.83 &  89.86   &  167.81 &    176.86 \\ % wrong number of solutions for clean
\hline
large\_curves   & 24, 19  & 137 & 110.20 & \textbf{91.15}            & 90.82                       & 376.71 & 377.55   &  501.52 &    138.35 \\
\hline 
degree\_6\_surf & 42      &  13 & 149.33 & \textbf{17.46}            & 16.50                       &  18.63 &  62.18   & timeout &    133.77\\
\hline
challenge\_12   & 40      &  99 & 108.74 & \textbf{23.07}            & 27.66                       &  27.20 & 195.76   &   41.13 &     40.86 \\
\hline
SA\_4\_4\_eps   & 33      &   2 & 155.92 &  \textbf{2.83}            &  2.85                       &   3.81 &   8.57   &  296.02 &     56.30 \\
\hline
FTT\_5\_4\_4    & 40      &  62 &  73.89 & \textbf{17.58}            & 20.92                       &  20.99 & 111.22   & timeout &    199.92 \\
\hline
dfold\_10\_6    & 32      &  21 &  26.20 &  \textbf{4.80}            &  3.12                       &   3.19 &   3.54   &    3.14 &      3.84 \\
\hline
cov\_sol\_20    & 20      &   8 &  27.25 & \textbf{12.36}            & 36.10                       &  42.81 &  52.16   &  762.80 &    175.85 \\
\hline
mignotte\_xy    & 32      &  30 & 545.88 &\textbf{438.38}            &440.64                       & 986.68 &1021.50   & timeout &   timeout\\
\hline
spider          & 28      &  38 & 389.06 & \textbf{81.63}            & 87.44                       & 135.15 & 314.56   & timeout &   timeout\\ % check number of solutions for clean
\hline
\end{tabular}
\footnotetext{\textbf{timeout:} algorithm timed out ($>$~1500~sec)}
\end{minipage}
\end{table}

\begin{table}[tb]
\centering\footnotesize\sffamily
\caption{Averaged running times for $10$ pairs of curves defined by random polynomials of degree~9 and 15 with increasing bit-lengths (given by {\bf shift} parameter). For description of configurations, see~Table~\ref{tbl:test1}.}
\label{tbl:test2}
\begin{minipage}[t]{\linewidth}
\centering
\begin{tabular}[]{|l|r|r| r||r | r||r | r|r ||r|r|} %p{2cm}|p{2cm}|p{2cm}|p{2cm}|p{2cm}|p{2cm}|}
\hline
Density of & & & avg. & \multicolumn{2}{|c||}{\textsc{BS+allfilters}} & \textsc{BS+bstr+comb} & \textsc{BS+bstr} & \textsc{BS} & \isolate & \lgp \\
 polynomials &$y$-degree & shift & \#sols & CPU & GPU & \multicolumn{3}{c||}{GPU} & Maple & Maple \\
\hline
\hline
\multirow{4}{*}{dense} & \multirow{4}{*}{9,9}      &  -   & \multirow{4}{*}{5.6} 
                                     &  0.30  & \textbf{0.21}                   & 0.22                    &  0.21  &   0.19  &   0.33 &   0.24   \\
   &  &  128   &                     &  0.48  & \textbf{0.15}                   & 0.16                    &  0.15  &   0.49  &   0.66 &   0.93   \\
   &  &  512   &                     &  2.22  & \textbf{0.31}                   & 0.31                    &  0.31  &   2.03  &   1.51 &   3.33   \\
   &  & 2048   &                     & 16.47  & \textbf{2.07}                   & 2.06                    &  2.07  &  13.86  &   7.48 & 102.37   \\
\hline
\multirow{4}{*}{dense} & \multirow{4}{*}{15,15} &  -   & \multirow{4}{*}{5.0} 
                                     &  1.82  & \textbf{0.71}                   & 0.70                    &  0.71  &   1.64  &   6.85 &   3.88   \\
   &  &  128   &                     &  6.02  & \textbf{0.69}                   & 0.67                    &  0.67  &   3.74  &  14.66 &   8.31   \\
   &  &  512   &                     & 32.18  & \textbf{1.48}                   & 1.45                    &  1.48  &  14.35  &  38.27 &  22.36   \\
   &  & 2048   &                     &251.07  & \textbf{8.97}                  & 8.94                    &  8.97  &  94.09  & 141.69 & 102.36   \\
\hline
\hline
\multirow{4}{*}{sparse} & \multirow{4}{*}{9,9}      &  -   & \multirow{4}{*}{4.5} 

                                     &  0.10  & \textbf{0.07}                   & 0.07                    &  0.07  &   0.09  &   0.07 &   0.22   \\
   &  &  128   &                     &  0.14  & \textbf{0.08}                   & 0.08                    &  0.07  &   0.21  &   0.20 &   0.57   \\
   &  &  512   &                     &  0.46  & \textbf{0.16}                   & 0.15                    &  0.15  &   0.85  &   0.70 &   1.74   \\
   &  & 2048   &                     &  3.11  & \textbf{0.84}                   & 0.84                    &  0.84  &   5.86  &   5.40 &   7.38   \\
\hline
\multirow{4}{*}{sparse} & \multirow{4}{*}{15,15} &  -   & \multirow{4}{*}{3.8} 
                                     &  0.65  & \textbf{0.36}                   & 0.36                    &  0.36  &   0.66  &   0.99  &  1.24   \\
   &  &  128   &                     &  1.55  & \textbf{0.46}                   & 0.47                    &  0.46  &   1.50  &   4.68  &  3.51   \\
   &  &  512   &                     &  7.70  & \textbf{1.55}                   & 1.55                    &  1.55  &   6.80  &  16.01  & 11.93   \\
   &  & 2048   &                     & 58.97  &\textbf{13.45}                   &13.38                    & 13.46  &  51.14  & 132.76  & 74.03   \\
\hline
\hline
\end{tabular}
\end{minipage}
\end{table}
In our experiments, we have used the number types provided by \gmp~4.3.1
and fast polynomial GCD from \ntl~5.5 
library.\footnote{\gmp:~\url{http://gmplib.org},
\ntl:~\url{http://www.shoup.net/ntl}}
All experiments have been run on 2.8GHz $8$-Core Intel Xeon W3530 with 8~MB
of L2~cache under Linux platform. For the GPU-part of the algorithm, we have
used the GeForce GTX480 graphics processor (Fermi Core).
We compared our approach to the bivariate version of
\isolate (based on \rs by 
Fabrice~Rouillier\footnote{\rs:~\url{http://www.loria.fr/equipes/vegas/rs}}) 
and \lgp by Xiao-Shan Gao~\etal\footnote{The software is available at
\url{http://www.mmrc.iss.ac.cn/~xgao/software.html}}
We remark that, for the important substep of isolating the real roots of the 
elimination polynomial, all three contestants (including our implementation) 
use the highly efficient implementation provided by \rs.

Our tests consist of two parts: In the first part, we consider ``special''
curves (and their derivative w.r.t. $y$-variable) selected in the aim of challenging different parts of the algorithm
and showing the efficiency of the filtering techniques given in
Section~\ref{ssec:filters}. These curves, for instance, have many 
singularities or high-curvature points which requires many candidates to 
be tested along each vertical line, or prohibit the use of special filters.
Descriptions of the considered curves and corresponding timings are listed in Table~\ref{tbl:test-set1} and Table~\ref{tbl:test1}, respectively. 
In the second part of our experiments, we study the performance of the
\bs on random polynomials with increasing total degrees and coefficient bit-lengths. We refer the reader to Table~\ref{tbl:test2} for the corresponding timings. 
Appendix~\ref{asec:benchmarks} features further experiments.

In columns $4$--$8$, the experiments for our algorithm are given with all 
filters set on (\textsc{BS+allfilters}), with bitstream and combinatorial
filter (\textsc{BS+bstr+comb}), with bitstream filter only
(\textsc{BS+bstr}) and with all filters set off 
(\textsc{BS}).
For \bs, we report timings respectively with and without GPU resultant algorithm. 
For the remaining configurations we show only the timings using GPU resultants. CPU-based timings can easly be obtained
by taking the difference between \bs-columns.

One can observe that our algorithm is generally 
superior to \isolate and \lgp even if the filters
are not used.
By comparing columns $5$--$8$ in the table, one can see that
filtering sometimes results in a
significant performance improvement. The \emph{combinatorial} test is
particularly useful when the defining polynomials of the system
(\ref{system})
have large degrees and/or large coefficient bit-length while at the same
time the number of
covertical or singular solutions is small compared to the total
number of candidates being checked.
The \emph{bidirectional} filter is advantageous when the system has
covertical solutions in one direction
(say along $y$-axis) which are \emph{not} cohorizontal.
This is essentially the case for
\textsf{challenge\_12}, \textsf{cov\_sol\_20} and \textsf{spider}.

Another strength of our approach relates to the fact that the amount of
symbolic operations is crucially reduced.
Hence, when the time for computing resultants is dominating, 
the GPU-based algorithm offers a speed-up by the factor of 
\textbf{$2$-$5$} over the version with default resultant implementation. 
It is also worth mentioning that both \isolate and \lgp benefit from 
the \emph{fast resultant computation} available in Maple~13 while 
\cgal's default resultant computation\footnote{Authors are 
indebted to \cgal developers working on resultants.} is generally 
\emph{much slower} than that of Maple. As a result, there is a large 
discrepancy columns~$4$ and~$5$ for \bs.

Table~\ref{tbl:test2} lists timings for experiments with random curves.
Each instance consists of five curves of the same degree 
(9 or 15, dense or sparse) and we report the average time to 
compute the solutions for one of all ten pairs of curves.
In order to analyze the influence of the coefficients' bit-lengths, 
we multiplied each curve by $2^k$ with $k \in \{128, 512, 2048\}$ and 
increased the constant coefficient by one.
Since the latter operation constitutes only a small 
perturbation of the vanishing set of the input system, the number of solutions 
remains constant while the content of the polynomials' 
coefficients also stays trivial. 
We see that the bidirectional filtering is not of any advantage because the
system defined by random polynomials is unlikely to have covertical
solutions. However, in this case, most candidates are rejected by the
combinatorial check, thereby omitting (a more expensive) test 
based on Theorem~\ref{thm:inclusion}. This results in a clear speed-up 
over a ``non-filtered'' version.
Also, observe that GPU-\bs is not vulnerable to increasing the 
bit-length of 
coefficients while this becomes critical for \isolate's and \lgp's 
performance. We have also observed that, for our filtered versions, the time 
for the validation step is almost independent 
of the bit-lengths.

We omit experiments to refine the solution boxes to certain precision
as this matches the efficiency of QIR due to the fact that we have 
algebraic descriptions for solutions' $x$- and $y$-coordinates.

\ignore{
\eric{existing: cgal, polynomial support~\cite{cgal:h-p-10}, real root 
solving (results in~\cite{HTZ-CROSSAK-2009}), bitstream, alg numbers 
(with quadratic interval refinement~\cite{abbott-qir-06})}

\eric{new: bi-algebraic-real, bisolve, certify using: interval arithmetic
for complex numbers, norm test}

\eric{arbitrary precision can be queried}

\subsection{Results}
\label{ssec:results}

We tested various families of curves ($f$ and $f_y$) and pair of curves.

\section{Summary and Outlook}
\label{sec:conclusion}
We propose an exact and complete method to isolate the real solutions
of a bivariate polynomial system. Our algorithm is designed to
reduce the number of purely symbolic operations as much as possible.
Eventually, only resultant computation and square-free factorization of
the resultant polynomial are still needed. By transferring the
resultant computation to the GPU, we are able to remove a major
bottleneck of elimination approaches. In order to further improve our 
implementation, we aim to outsource the square-free factorization to the 
GPU as well, a step which seems to be feasible since factorization is also
well suited for a "divide-conquer-combine" modular approach.
Since our initial motivation was to speed up the topology and arrangement
computation for algebraic curves and surfaces, we plan to extend our method
towards this direction. Furthermore, it would be interesting to extend 
our algorithm to handle higher dimensional systems or complex solutions.
Finally, we would like to investigate in hybrid methods such as the 
combination of a numerical complex root solver and an exact post 
certification method serving as an additional filter in the validation step (in
the spirit of~\cite{strzebonski-cylindrical,r-tenmethods}). We are 
convinced that most of the candidate boxes could be treated even more 
efficiently by 
the use of such methods. We claim that, eventually, the  
total costs for solving a bivariate system should \emph{only} be dominated by 
those of the root isolation step for the elimination polynomial. For many 
instances, our experiments already hint to the latter claim. We aim to further 
improve our implementation to show this behavior for all instances and to 
provide a proof in terms of complexity as well.

We have seen promising results 

A direct extension is to consider the topologocial analysis of a single 
curve (or pair or curves). Beyond, there are various geometric applications
 supported.

parameter space of quadrics with non-integer 
coefficients~\cite{dupont07abc}

One of our goals to investigate further is the three- or even multivariate 
case.
}

\section{Summary and Outlook}
\label{sec:conclusion}
We propose an exact and complete method to isolate the real solutions
of a bivariate polynomial system. Our algorithm is designed to
reduce the number of purely symbolic operations as much as possible.
Eventually, only resultant computation and square-free factorization of
the resultant polynomial are still needed. By transferring the
resultant computation to the GPU, we are able to remove a major
bottleneck of elimination approaches. In order to further improve our 
implementation, we aim to outsource the square-free factorization to the 
GPU as well, a step which seems to be feasible since factorization is also
well suited for a "divide-conquer-combine" modular approach.
Since our initial motivation was to speed up the topology and arrangement
computation for algebraic curves and surfaces, we plan to extend our method
towards this direction. Furthermore, it would be interesting to extend 
our algorithm to handle higher dimensional systems or complex solutions.
Finally, we would like to investigate in hybrid methods such as the 
combination of a numerical complex root solver and an exact post 
certification method serving as an additional filter in the validation step (in
the spirit of~\cite{strzebonski-cylindrical,r-tenmethods}). We are 
convinced that most of the candidate boxes could be treated even more 
efficiently by 
the use of such methods. We claim that, eventually, the  
total costs for solving a bivariate system should \emph{only} be dominated by 
those of the root isolation step for the elimination polynomial. For many 
instances, our experiments already hint to the latter claim. We aim to further 
improve our implementation to show this behavior for all instances and to 
provide a proof in terms of complexity as well.

{\newpage
\bibliographystyle{abbrv}
\bibliography{bib,bib2}
}

\newpage
\begin{appendix}

\section{Proofs}
\label{asec:proofs}
\begin{theorem}
\label{thm:testappendix}
Consider a disk $\Delta=\Delta_m(r)\subset\C$ with center $m$ and radius
$r$.
\begin{enumerate}
\item If $T_{K}^p(m,r)$ holds for some $K\geq 1$, then the closure
$\overline{\Delta}$ of $\Delta$ contains no root of $p$.
\item	If $T^{p'}_{K}(m,r)$ holds for a $K\geq \sqrt{2}$, then
$\overline{\Delta}$ contains at most one root of $p$.
\end{enumerate}
\end{theorem}

\begin{proof}
(1) follows from a straight forward computation: For each
$z\in\overline{\Delta}$, we have $$p(z)=p(m+(z-m))=p(m)+\sum_{k\ge 1}
		\frac{p^{(k)}(m)}{k!}(z-m)^k$$ and, thus,
		$$\frac{|p(z)|}{|p(m)|}\geq 1-\frac{1}{|p(m)|}\cdot\sum_{k\ge 1}
		\frac{|p^{(k)}(m)|}{k!}|z-m|^k>\left(1-\frac{1}{K}\right)$$
since $|z-m|\le r$ and $T_{K}^p(m,r)$ holds. In particular, for $K\geq 1$,
the above inequality implies $|p(z)|>0$ and, thus, $p$ has no root in
$\overline{\Delta}$.

It remains to show (2): If $T_K^{p'}(m,r)$ holds, then, for any point
$z\in\overline{\Delta}$, the derivative $p'(z)$ differs from $p'(m)$ by a
complex number of absolute value less than $|p'(m)|/K$. Consider the
triangle spanned by the points $0$, $p'(m)$ and $p'(z)$, and let $\alpha$
and $\beta$ denote the angles at the points $0$ and $p'(z)$, respectively.
From the Sine Theorem, it follows that $$|\sin\alpha|=|p'(m)-p'(z)|\cdot
\frac{|\sin \gamma|}{|p'(m)|}< \frac{1}{K}.$$ Thus, the arguments of
$p'(m)$ and $p'(z)$ differ by less than $\operatorname*{arcsin}(1/K)$ which
is smaller than or equal to $\pi/4$ for $K\geq\sqrt{2}$.
Assume that there exist two roots $a, b\in\overline{\Delta}$ of $p$. Since
$a=b$ implies $p'(a)=0$, which is not possible as $T^{p'}_1(m,r)$ holds, we
can assume that $a\neq b$.
We split $f$ into its real and imaginary part, that is, we consider
$p(x+iy)=u(x,y)+iv(x,y)$ where $u,v:\RR^2\rightarrow\RR$ are two bivariate
polynomials. Then, $p(a)=p(b)=0$ and so $u(a)=v(a)=u(b)=v(b)=0$.
But $u(a)=u(b)=0$ implies, due to the Mean Value Theorem in several real
variables~\cite{},
that there exists a $\phi\in [a,b]$ such that 
	$$\nabla u(\phi)\perp (b-a).$$
Similarly, $v(a)=v(b)=0$ implies that there exists a $\xi\in [a,b]$
such that $\nabla v(\xi)\perp (b-a).$ But $\nabla
v(\xi)=(v_x(\xi),v_y(\xi))=(-u_y(\xi), u_x(\xi))$, thus,
it follows that $\nabla u(\xi) ~\|~ (b-a).$
Therefore, $\nabla u(\psi)$ and $\nabla u(\xi)$ must be perpendicular.
Since $p'=u_x+iv_x=u_x-iu_y$, the arguments of $p'(\psi)$ and $p'(\xi)$
must differ by $\pi/2$. This contradicts our above result that both differ
from the argument of $p'(m)$ by less than $\pi/4$, thus, (2) follows.
\end{proof}

\newpage
\section{Further Experiments}
\label{asec:benchmarks}

\begin{table}[ht]
\centering\footnotesize\sffamily
\caption{Description of the curves used in experiments. In case only
a single curve is given, the second curve is taken to be the first
derivative w.r.t.\ $y$-variable.}
\label{tbl:app-test-set1}
\begin{minipage}[t]{\linewidth}
\centering
 \renewcommand{\thefootnote}{*}  
\begin{tabular}[]{|l|p{2.5in}||l|p{2.5in}|}
\hline
Instance  & Description & Instance  & Description \\[3pt]
\hline\hline
hard\_one &  vertical lines as component of one curve, many candidates to test & compact\_surf  & silhouette of an algebraic surface, many singularities, isolated solutions  \\
\hline
grid\_deg\_10  & large coefficients, curve in generic position & 13\_sings\_9 & large coefficients, high-curvature points  \\
\hline
huge\_cusp  & large coefficients, high-curvature points & swinnerston\_dyer  & covertical solutions in both directions \\
\hline
cusps\_and\_flexes  & high-curvature points & challenge\_12\_1\footnotemark & many candidate solutions to be checked  \\
\hline
L6\_circles  & $4$ circles w.r.t.\ L6-norm, clustered solutions & SA\_2\_4\_eps\footnotemark  & singular points with high tangencies, displaced  \\
\hline
ten\_circles  & set of $10$ random circles multiplied together, rational solutions &
 spiral29\_24 & taylor expansion of a spiral intersecting a curve with many branches, many candidates to check \\
\hline
curve24 & curvature of degree 8 curve, many singularities & & \\
\hline
\end{tabular}
\footnotetext{* These curves were taken from~\cite{labs_10}}
\addtocounter{footnote}{-2}
\end{minipage}
\end{table}

\begin{table}[ht]
\centering\footnotesize\sffamily
\caption{Results for the curves listed in Table~\ref{tbl:app-test-set1}.
We used the same configurations as in Table~\ref{tbl:test1}.}
\label{tbl:app-test1}
\begin{minipage}[t]{\linewidth}
\centering
\begin{tabular}[]{|l| r| r||r | r||r | r|r ||r|r|} %p{2cm}|p{2cm}|p{2cm}|p{2cm}|p{2cm}|p{2cm}|}
\hline
 & & & \multicolumn{2}{|c||}{\textsc{BS+allfilters}} & \textsc{BS+bstr+comb} & \textsc{BS+bstr} & \textsc{BS} & \isolate & \lgp \\
Instance &$y$-degree & \#sols & CPU & GPU & \multicolumn{3}{c||}{GPU} & Maple & Maple \\
\hline
\hline
hard\_one          & 27, 6  & 46 &  8.17  &   \textbf{6.95}               &  6.96                        &  12.44   &  12.09  &   25.20 &  20.00 \\
\hline
grid\_deg\_10      & 10     & 20 &  4.05  &   \textbf{1.63}               &  1.64                        &   3.22   &   3.01  &  106.95 &   3.16 \\
\hline
huge\_cusp         & 8      & 24 & 33.24  &  \textbf{21.43}               & 21.15                        &  26.97   &  26.47  &  768.56 & 119.03 \\
\hline
cusps\_and\_flexes & 9      & 20 &  2.31  &   \textbf{1.37}               &  1.28                        &   1.70   &   1.38  &   28.42 &   2.73 \\
\hline
L6\_circles        & 24     & 18 & 25.00  &   \textbf{6.21}               &  5.68                        &   6.88   &   5.08  &   46.61 &  52.79 \\
\hline
curve24            & 24     & 28 & 41.26  &  \textbf{16.61}               & 16.66                        &  118.91  & 115.62  &   49.69 &  41.96 \\
\hline
ten\_circles       & 20     & 45 & 10.51  &   \textbf{7.64}               &  4.63                        &    4.93  &   2.57  &    5.22 &   5.24 \\
\hline
compact\_surf      & 18     & 57 & 19.01  &   \textbf{6.53}               &  5.98                        &    5.85  &  23.56  & timeout &  12.31 \\
\hline
13\_sings\_9       & 9      & 35 &  3.38  &   \textbf{2.39}               &  2.23                        &    2.98  &   2.41  &   28.60 &   2.97 \\
\hline
swinnerston\_dyer  & 40     & 63 & 50.32  &  \textbf{22.60}               & 22.53                        &   22.33  &  56.46  &   71.00 &  28.47 \\
\hline
challenge\_12\_1   & 30     & 99 & 24.67  &   \textbf{9.17}               &  9.89                        &    9.44  &  41.84  &   41.13 &  40.86 \\
\hline
SA\_2\_4\_eps      & 17     &  6 &  6.71  &   \textbf{0.56}               &  0.59                        &    0.70  &   1.66  &    7.83 &   4.90 \\
\hline
spiral29\_24       & 29, 24 & 51 & 80.37  &  \textbf{35.34}               & 35.13                        &  290.35  & 286.79  &  144.79 &  84.97 \\
\hline
\end{tabular}
\footnotetext{\textbf{timeout:} algorithm timed out ($>$~1500~sec)}
\end{minipage}
\end{table}

\end{appendix}

\end{document}